\documentclass[12pt]{article}
\usepackage{graphicx,color}
\usepackage[utf8]{inputenc}
\usepackage{amssymb,amsmath,amsthm,amsfonts}

\def\RR{{\mathbb R}}
\def\CC{{\mathbb C}}

\def \eins {{\mathbf 1}}
\def\eps{\varepsilon}

\def\sl2{{{\rm SL}(2,\RR)}}
\def\psl2{{{\rm PSL}(2,\RR)}}
\def\u1{{{\rm V}(1)}}
\def\su2{{{\rm SV}(2)}}

\def\so3{{{\rm SO}(3)}}

\def\Mob{{\rm\textsf{M\"ob}}}

\def\SO{{\mathrm{SO}}}

\def\ket#1{\vert#1\rangle}
\def\bra#1{\langle#1\vert}
\def\kets#1{\vert#1\rangle^{\rm scl}}
\def\ketm#1{\vert#1\rangle^{\rm Maxw}}
\def\Ket#1{\big\vert#1\big\rangle}
\def\KET#1{\Big\vert#1\Big\rangle}
\def\braket#1#2{\langle#1\vert#2\rangle}

\def\wh{\widehat}
\def\ol{\overline}

\def\PP{{\widetilde P}}
\def\MM{{\widetilde M}}
\def\KK{{\widetilde K}}
\def\HH{{\mathcal H}}
\def\HHs{{\mathcal H}_{\rm scl}}
\def\HHm{{\mathcal H}_{\rm Maxw}}
\def\FF{{\mathcal F}}
\def\DD{{\mathcal D}}
\def\AA{{\widetilde A}}
\def\UU{{\widetilde U}}

\def\cf{{\mathfrak{so}(2,4)}}

\def\rr{{\mathfrak{so}(3)}}
\def\t{{\mathfrak{t}}}
\def\ad{{\operatorname{ad}}}
\def\Ad{{\operatorname{Ad}}}
\def\mb{{\mathfrak{m\ddot ob}}}
\def\hh{{\mathfrak{h}}}
\def\mm{{\mathfrak{m}}}

\def\eref#1{(\ref{#1})}
\def\pref#1{Prop.\ \ref{#1}}
\def\sref#1{Sect.\ \ref{#1}}
\def\bea#1{\begin{eqnarray}\label{#1}}
\def\eea{\end{eqnarray}}
\def\ba{\begin{array}}
\def\ea{\end{array}}

\newtheorem{theorem}{Theorem}[section]
\newtheorem{lemma}[theorem]{Lemma}

\newtheorem{proposition}[theorem]{Proposition}

\title{\Large\bf \vskip-2cm Spacelike deformations: \\ Higher-helicity fields
  from scalar fields }

\author{{\sc 
Vincenzo Morinelli} \footnote{Titolare di un Assegno di Ricerca
dell'Istituto Nazionale di Alta Matematica (INdAM fellowship).} 
\\
Dipartimento di Matematica,
Universit\`a di Roma Tor Vergata,\\
Via della Ricerca Scientifica, 1, I-00133 Roma, Italy\\
E-mail: {\tt 
morinell@mat.uniroma2.it}
\\[5mm]
{\sc Karl-Henning Rehren} 
\\
Institut f\"ur Theoretische Physik, Universit\"at G\"ottingen,\\ 
37077 G\"ottingen, Germany\\
E-mail: {\tt rehren@theorie.physik.uni-goettingen.de}
}

\begin{document}

\maketitle

\begin{abstract}
  {In contrast to Hamiltonian perturbation theory which
  changes the time evolution, ``spacelike deformations'' proceed by
  changing the translations (momentum operators). The free Maxwell
  theory is only the first member of an infinite family of spacelike
  deformations of the complex massless Klein-Gordon quantum field into
  fields of higher helicity. A similar but simpler
    instance of  spacelike deformation allows to increase the mass of
    scalar fields.}
\end{abstract}
  \parindent0mm
\parskip2mm

\section{Introduction}
The basic idea of Hamiltonian perturbation theory is to start from
a time zero algebra (``canonical commutation relations'') equipped
with a free time evolution, and perturb the free Hamiltonian such that
the observables at later time $\Phi(t):=e^{iHt}\Phi_0e^{-iHt}$ (where
$H$ is the perturbed Hamiltonian) deviate from the free ones. 
We present here a ``complementary'' deformation scheme for free
quantum field theories: 
fixing the algebra along the time axis, we deform the space
translations, so as to obtain a different local quantum field theory in
Minkowski space.

Despite the apparent similarity, there are many differences,
though. Hamiltonian Perturbation Theory (PT) is well-known to be
obstructed by Haag's theorem, which implies that the perturbation is
possible on the same 
Hilbert space only locally. Globally, the perturbed vacuum state is
not a state in the ``free Hilbert space'', so that one is forced to
change the representation of the time zero algebra. The need of
renormalization of the mass also shows that one is even forced to
change the time zero algebra itself. More precisely, interacting
quantum fields in general do not even exist as distributions at a fixed time (see, e.g., \cite{S,W}).

A recent approach \cite{BF}, designed to avoid these obstructions, uses instead
of a CCR time-zero algebra, an abstract ``off-shell'' C*-algebra of
kinematical fields on spacetime which supports a large class of
dynamics (one-parameter groups of time-evolution automorphisms). The
invariant states under each dynamics, however, annihilate different
ideals of the algebra (``field equations''), such that the
corresponding GNS Hilbert spaces cannot be identified for any
time-zero subalgebra. 

In contrast, Wightman quantum fields can always be restricted to the
time axis \cite{Bo}. Our spacelike deformations are globally
well-defined on a subspace of the original Hilbert space. They consist in a
redefinition of the generators of the spacelike translations (momentum
operators). The perturbed fields away from the time axis are then
defined as $\Phi(t,\vec x)=e^{i x^k\PP_k}\Phi_0(t,0)
e^{-i  x^k\PP_k}$, where $\PP_k$ are the deformed generators. 

In Hamiltonian PT, the ``field content'' is fixed by the choice of the
free theory. The ``particle content'' is determined by the spectrum of
the (renormalized) perturbed Hamiltonian, and may well change, e.g.,
when the interacting theory has bound states, or confinement
occurs. Yet, the relation to the free particle content is usually not
entirely lost.

In contrast, our spacelike deformations are 
(non-perturbative) algebraic deformations that drastically change the field
content without changing the Hamiltonian: e.g., one obtains the free
Maxwell field by a deformation of a massless free scalar field.  

In fact, we know spacelike deformations only for free fields,
producing massless higher helicity fields from scalar ones (\sref{s:main}), or
massive scalar fields from massless ones (\sref{s:mass}). The reason is that (a) we work on
the one-particle space, from which the deformation passes to the Fock
space by ``second quantization''; and (b)
the construction is essentially representation-theoretic. Namely,
  it transfers the representation of the Poincar\'e group of one
  theory to a subspace of the representation space of the other
  theory by a unitary operator, intertwining the subgroup that
  fixes the time-axis.\footnote{This subgroup consists of the rotations and the
    M\"obius group $\Mob$. The latter is familiar from
    chiral conformal QFT on the light-ray of two-dimensional
    CFT. Here, $\Mob$ acts in the same way by 
  fractional linear transformations on the (compactified) time
  axis. It is generated by
  the time translations and the ``conformal inversion'' $x=(t,\vec
  x)\mapsto \frac{(-t,\vec x)}{t^2-\vec x\,^2}$, which becomes
  $t\mapsto -1/t$ on the time axis.} We present the  deformed generators
as (nonlinear) functions of the undeformed generators.

We are therefore far from ``interactions via deformation'';  
but our models illustrate the potential of a new approach, and more
sophisticated new ideas may emerge from the present simple
prototypes. 

\section{Helicity deformations}
\setcounter{equation}{0}
\label{s:heli}
\subsection{Background}
\label{s:back}

The first examples to be demonstrated rely on a recent observation in \cite{LMPR}:
For the massless free fields of any integer helicity $h>0$, the one-particle
spaces $\HH^{(h)}$ are proper subspaces of the one-particle space $\HH=\HH^{(0)}$ of
the complex massless free scalar field. More precisely, $\HH^{(h)}$
as representations $U_{h}\oplus U_{-h}$ of the
Poincar\'e group extend to representations $U_{(h)}$ of the conformal group,
whose restriction to the subgroup $\Mob \times \SO(3)$ is given by 
\bea{repns}
\HH^{(h)}\vert_{\Mob\times\SO(3)}=\bigoplus_{\ell=h}^\infty
(U^{(\ell+1)}\otimes \DD^{(\ell)})\oplus (U^{(\ell+1)}\otimes
\DD^{(\ell)}),
\eea
where $\DD^{(\ell)}$ are the spin-$\ell$ representations of $\SO(3)$,
and $U^{(d)}$ are the irreducible positive-energy representations of
$\Mob$ with lowest eigenvalue $d$ of the ``conformal Hamiltonian''
$L_0=\frac12(P_0+K_0)$. (We follow rather standard notation for the
generators of the conformal group, fixing conventions in \sref{s:prel}; for more details on the relations
between the various groups and representations, see \sref{s:prel}
and \cite{LMPR}.)
The doubling is due to the
``electric'' and ``magnetic'' degrees of freedom. The same
decomposition with $h=0$ holds for the complex scalar field, where the
doubling corresponds to the subspaces of charge $\pm 1$. 

$\Mob \times \SO(3)$ is the
subgroup of the conformal group that fixes the time axis $\vec
x=0$. The vectors transforming in the displayed subrepresentations are
spacelike derivatives of fields on the time axis, that
transform like quasiprimary fields under $\Mob$, applied to the vacuum
vector $\Omega$. For the scalar field, these fields are simply \cite{BDL}
$$Y_\ell(\vec\nabla)\varphi^{(*)}(x)\vert_{x=(t,\vec 0)}$$ 
($\varphi^{(*)}$ stands for both $\varphi$ and the
conjugate field $\varphi^*$) with harmonic\footnote{A polynomial in $\RR^n$ is
  harmonic iff $\Delta Y=0$ where $\Delta=\sum_{k=1}^n\partial_k^2$ is the
Laplacian. In this paper, $n=3$.}  homogeneous
polynomials $Y_\ell$ of degree $\ell$, transforming like spin-$\ell$
multiplets of quasiprimary
fields of scaling dimension $d=\ell+1$. For $h>0$, 
when the electric and magnetic fields are combined into a complex
field tensor $F_{j_1\dots j_h}=E_{j_1\dots j_h}+iB_{j_1\dots j_h}$,
the equations of motion impose linear relations among the fields
$\nabla_{i_1}\dots\nabla_{i_r} F_{j_1\dots j_h}(x)\vert_{x=(t,\vec 0)}$.
The decomposition \eref{repns} implies that the time-axis field content is
given by exactly two quasiprimary
spin-$\ell$ multiplets (one for both $F$ and $F^*$) of scaling dimension
$d=\ell+1$ for each $\ell\geq h$, and that the Casimir operator of the
Lie algebra of the rotations (eigenvalues $\ell(\ell+1)$) and the
Casimir operator of the Lie algebra of the M\"obius group 
(eigenvalues $d(d-1)$) coincide in the one-particle representation.

In this count, as $h$ increases, the
field content decreases. The lowest fields of
the scalar theory are given by $\varphi^{(*)}(t,\vec 0)$ ($\ell=0$, $d=1$) and 
$\vec \nabla \varphi^{(*)}(t,\vec 0)$ ($\ell=1$, $d=2$), while, e.g., the Maxwell
theory starts at $\ell=1$, $d=2$ with the vector field $\vec F^{(*)}=\vec
E\pm i\vec B$. In this sense, contrary to intuition, the higher-helicity
theories have {\em less} degrees of freedom than the lower-$h$
theories. 

In \cite{LMPR}, these facts were exploited to estimate the trace of
$e^{-\beta L_0}$, whose finiteness then implies the split property for
all finite-helicity massles free quantum field theories. Here, we take
them as the starting point of spacelike deformation, as already
speculated in \cite{LMPR}: we unitarily identify the
common subrepresentations $\ell\geq h$ of the subgroup
$\Mob\times\SO(3)$, and find the necessary modification of the
representation of the remaining generators. 

To illustrate the idea, consider the
case $h=1$ (Maxwell). The Maxwell equations for $\vec F$ read
$$\vec\nabla\cdot\vec F=0,\quad \vec\nabla\times\vec F = i\partial_t
\vec F.$$ 
The component fields $F_k(t,\vec 0)$ on the time axis transform in the
same way under M\"obius transformations of the time axis and
rotations, like the fields $\nabla_k\varphi(x,\vec 0)$ of the
complex massless Klein-Gordon theory. Similarly, the fields
$(\nabla_iF_j+\nabla_jF_i)(t,\vec0)$ of spin 2 transform in the same way as the fields
$(\nabla_i\nabla_j-\frac13\delta_{ij}\Delta)\varphi(t,\vec 0)$. 

Because the representations of the M\"obius and rotation groups on the
one-particle spaces are the same -- except for the absence of the
subrepresentation with $\ell=0$ in the Maxwell theory -- we can
algebraically identify these pairs of fields along the time axis. We get 
\bea{Id1} F_i(t,\vec 0) \,\stackrel!=\, 2\nabla_i\varphi(t,\vec 0)\eea
\bea{Id2} \nabla_iF_j(t,\vec0) \,\stackrel!=\, \alpha\cdot
\Big(\nabla_i\nabla_j-\frac13\delta_{ij}\Delta\Big)\varphi(t,\vec 0)
+i\eps_{ijk}\partial_t\nabla_k\varphi(t,\vec 0),\eea
where the Maxwell equations dictate the anti-symmetric part in
\eref{Id2} as well as the absence of 
an $\ell=0$ contribution; the
two-point function fixes the normalizations, giving $\vert\alpha\vert^2=12$.

The problem is apparent: the left-hand side of \eref{Id2} is the
derivative of the left-hand side of \eref{Id1}, which is not true for
the right-hand sides. The spatial derivatives being implemented by the
momentum operators $P_k$, we conclude that while the M\"obius and
rotation generators of both theories (including the Hamiltonian $P_0$)
can be identified, their spatial momentum operators must differ. 

We are going to determine the momentum operators $\PP_k$ of the Maxwell theory
as polynomials of the conformal and charge generators of the Klein-Gordon
theory (and along with them the boosts and the generators of spatial special
conformal transformations). Then, starting from the identification
\eref{Id1} as a definition of the Maxwell field on the time axis, and
acting with $\UU(\vec x)=e^{i  x^k\PP_k}$ on $\varphi(t,\vec 0)$, one
obtains the Maxwell field everywhere in Minkowski space. The same
works for any helicity $h>0$. 

As a second instance, we present the spacelike deformation of the massless
scalar field into the massive scalar field in \sref{s:mass}. 

The mere existence of such deformations should not be too surprising,
given that ``all Hilbert spaces are the same''. The noticeable facts
are that the deformations fix parts of the symmetry, and that they can
be given on the remaining generators by explicit
formulae. 

\subsection{Preliminaries about the conformal Lie algebra}
\label{s:prel}

We denote by $P_\mu$, $M_{\mu\nu}$, $D$, $K_\mu$ the generators of
translations, Lorentz transformations, dilations, and special
conformal transformations in the conformal Lie algebra $\cf$,
respectively. Their commutators are explicitly 
\bea{cfLA}\ba{lll}
i[P_\mu,P_\nu]=0, & i[P_\mu,M_{\kappa\lambda}] =
\eta_{\mu\lambda} P_\kappa -\eta_{\mu\kappa} P_\lambda, & i[M_{\kappa\lambda},M_{\mu\nu}] =  \eta_{\kappa\mu} M_{\lambda\nu}
\pm \dots; \\[1mm] 
i[D,P_\mu]=P_\mu,& i[D,K_\mu] = -K_\mu, & i[D,M_{\kappa\lambda}]=0; \\[1mm] 
i[K_\mu,K_\nu]=0, & i[M_{\kappa\lambda},K_\mu] = \eta_{\kappa\mu}
K_\lambda-\eta_{\lambda\mu} K_\kappa,&
i[P_\mu,K_\nu]=-2\eta_{\mu\nu} D +2 M_{\mu\nu}.
\ea\eea
In particular, we have the Lie subalgebras $\mb$:
$$i[D,P_0]=P_0,\quad i[P_0,K_0]= -2D,\quad i[D,K_0]=-K_0, $$
and $\rr$:
$$i[M_{ij},M_{kl}] = \delta_{jk} M_{il} - \delta_{jl} M_{ik} -
\delta_{ik} M_{jl} + \delta_{il} M_{jk}.$$

\begin{lemma}
The parity reflection $(t,\vec x)\mapsto (t,-\vec x)$ defines a
symmetric space decomposition of the conformal Lie algebra
$$\cf = \hh\oplus\mm,\quad [\hh,\hh]\subset\hh\quad
[\hh,\mm]\subset\mm\quad [\mm,\mm]\subset\hh,$$
where $\hh=\mb\oplus\rr =\mathrm{Span}(P_0,D,K_0,M_{kl})$ and $\mm=\mathrm{Span}(P_k,M_{0k},K_k)$. The
generators of $\mm$ transform like vectors under $\rr$:
\bea{vector} i[M_{kl},X_i]=\delta_{li}X_k-\delta_{ki}X_l,\eea
and $\mb$ acts on $\mm$ as a $\mb$-module like
\bea{module}
\ba{lll}i[P_0,P_k]=0, & i[P_0,M_{0k}]=-P_k, &
i[P_0,K_k]=2M_{0k},
\\[1mm] 
i[D,P_k]=P_k, & i[D,M_{0k}]=0, & i[D,K_k]=-K_k,
\\[1mm] 
i[K_0,P_k]=2M_{0k}, & i[K_0,M_{0k}]=-K_k, & i[K_0,K_k]=0.\ea
\eea
In particular, $(\ad_{P_0})^3 = 0$ and $(\ad_{K_0})^3=0$ on $\mm$. 
\end{lemma}

Proof: Immediate from \eref{cfLA}. \qed

In the sequel, we shall need the action of the generators on
  suitable vectors in the $h=0$ representation. This
  representation is realized on the one-particle space $\HH$ of the
  real massless
  scalar field $\varphi$. Its two-point function
  $$(\varphi(x)\Omega,\varphi(x')\Omega) = (2\pi)^{-3}\int \frac
  {d^3p}{2p^0}\,e^{-ip(x-y)} \qquad (p^0=\vert \vec p\vert)$$
  is invariant under the infinitesimal conformal
  transformations\footnote{We use the same symbol for the
    second-quantized generators on the Fock space, as for the
    generators on the one-particle space.}
\bea{CR}
\ba{ll}i[P_\mu,\varphi(x)]=\partial_\mu\varphi(x),&
i[M_{\mu\nu},\varphi(x)]=(x_\mu\partial_\nu
-x_\nu\partial_\mu)\varphi(x),
\\[1mm] 
i[D,\varphi(x)]=((x\partial)+1)\varphi(x),&
i[K_\mu,\varphi(x)]=(2x_\mu(x\partial)-x^2\partial_\mu +
2x_\mu)\varphi(x).\ea
\eea
Because Wightman fields can be restricted to the time axis \cite{Bo},
they only need smearing in the time variable. We may thus introduce, for
polynomials $Y(\vec p)$ on momentum space $\RR^3$, improper one-particle vectors
\bea{IVec}\ket
Y_t =Y(-i\vec\nabla)\varphi(\vec x,t)\vert_{\vec
  x=0}\,\Omega,
\eea
such that $\ket{Y,f}:= \int f(t)\,\ket{Y}_t\,dt$ are proper
one-particle vectors. They span the one-particle space \cite{BDL}.
Their inner product is,  in spherical coordinates, 
\bea{IPY}\braket{Y,f}{Y',f'}=(2\pi)^{-3}\int_0^\infty
\frac{p^2\,dp}{2p}\, \ol{\wh f(p)}\wh f'(p)\cdot \int d\sigma(\vec n)\, \ol{Y(p\vec
  n)}Y'(p\vec n),
\eea
where
$p=\vert\vec p\vert$, $\vec p=p\vec n$, and $d\sigma$ is the invariant
measure on the unit sphere. The wave equation $\square \varphi=0$ states
that 
\bea{waveY}\ket{\vec p\,^2\,Y}_t+\partial_t^2\ket{Y}_t =0
\quad\Leftrightarrow\quad
\ket{\vec p\,^2\,Y,f}+\ket{Y,\partial_t^2f}=0\eea
i.e., these vectors have zero norm w.r.t.\ the inner product \eref{IPY}.

$\SO(3)$ acts on $Y(\vec p)$ by rotation of the
argument. The harmonic homogeneous polynomials $Y_\ell$ of degree $\ell$ carry
the irreducible representation $\DD^{(\ell)}$. Hence $\ket{Y_\ell}_t$
belong to $E_\ell\HH$.

The
inner product \eref{IPY} for harmonic homogeneous polynomials is diagonal w.r.t.\
$\ell$, and the resulting $p$-integral in \eref{IPY}
\bea{curr}
\int_0^\infty p^{2\ell+1}\,dp\, \ol{\wh f(p)}\wh f'(p)
\eea
is the two-point function of a chiral quasi-primary field (``conformal
current'') of scaling
dimension $d=\ell+1$,
which defines the inner product for the representation $U^{(\ell+1)}$ of $\Mob$.
In this way, the decomposition \eref{repns} for $h=0$ becomes manifest
\cite{BDL}. 

From \eref{CR}
and the invariance of the vacuum vector $\Omega$, one reads off 
the action of the conformal generators on $\ket{Y_\ell}_t$
for harmonic homogeneous polynomials $Y_\ell$ of degree $\ell$. They act 
(for simplicity of notation on the improper states)
as differential operators on $Y$ and w.r.t.\ $t$:
\bea{actmb} P_0\ket{Y_\ell}_t &=& -i\partial_t\ket{Y_\ell}_t,
\nonumber \\
D\ket{Y_\ell}_t &=& -i(t\partial_t+\ell+1)\ket{Y_\ell}_t,\\
K_0\ket{Y_\ell}_t &=& -i(t^2\partial_t+2(\ell+1)t\ket{Y_\ell}_t
\nonumber\eea
\bea{actso} M_{kl}\ket{Y_\ell}_t =
-i\Ket{(p_k\partial_l-p_l\partial_k) Y_\ell}_t\eea
\bea{actm}P_k \ket{Y_\ell}_t &=& \ket{p_kY_\ell}_t,\nonumber \\
\qquad M_{0k} \ket{Y_\ell}_t &=&
t\ket{p_kY_\ell}_t-\partial_t\ket{\partial_kY_\ell}_t,\\
K_{k} \ket{Y_\ell}_t &=&
-t^2\ket{p_kY_\ell}_t+2(t\partial_t+\ell)\ket{\partial_k Y_\ell}_t.
\nonumber \eea
The equality of the Casimir operators $\frac12 M_{kl}M_{kl}$ of $\rr$
and $\frac12(P_0K_0+K_0P_0)-D^2$ of $\mb$ with eigenvalues
$\ell(\ell+1)$ on $\ket{Y_\ell}_t$, as well as the mass-shell
condition $P_k^2=P_0^2$ can be directly
verified from these formulae and \eref{waveY}.

We shall need to control the spin of the vectors on the
  right-hand sides. The generators of $\rr$ and of $\mb$ clearly commute with
  $E_\ell$, hence \eref{actmb} and \eref{actso} have again spin $\ell$. 
  The polynomial $\partial_k Y_\ell$ is again
harmonic, hence $\ket{\partial_k Y_\ell}_t$ in \eref{actm} has spin
$\ell-1$. Since $p_kY_\ell-\frac{\vec
    p\,^2}{2\ell+1}\partial_kY_\ell$ is harmonic, the vector $\ket{p_kY_\ell}_t$
  splits into
  \bea{deco-pY}
 E_{\ell+1}\ket{p_kY_\ell}_t = \KET{\Big(p_k-\frac{\vec
    p\,^2}{2\ell+1}\partial_k\Big)Y_\ell}_t \quad\hbox{and}\quad
E_{\ell-1}\ket{p_kY_\ell}_t = -\frac1{2\ell+1}\partial_t^2
\ket{\partial_k Y_\ell}_t,\eea
where we have used \eref{waveY} in the latter. All other projections vanish.

Finally, for the complex scalar field, we have two copies of states
$\ket Y^\pm_t$ with the same actions \eref{actmb}--\eref{actm} of $\cf$ and actions of the
unitary charge and anti-unitary PCT operators
\bea{QJ}Q\ket Y^\pm_t= \pm \ket Y^\pm_t,\qquad J\ket Y^\pm_t = \ket
{Y}^\mp_{-t}.
\eea
$Q$ commutes with the conformal generators, while
\bea{PCT}
JP_\mu = P_\mu J,\quad JM_{\mu\nu} = -M_{\mu\nu}J, \quad
JD=-DJ,\quad JK_\mu = K_\mu J, \quad J Q = -Q J.
\eea

\subsection{Main result} 
\label{s:main}

Let $\HH=\HH^+\oplus \HH^-$ the one-particle space of the complex massless
Klein-Gordon field, where the superscript $\pm$ stands for the
eigenvalue $\pm 1$ of the charge operator $Q$. As representations of
$\Mob\times\SO(3)$, both $\HH^\pm$ decompose as
\bea{dec0}
\HH^\pm\vert_{\Mob\times\SO(3)}=\bigoplus_{\ell=0}^\infty \HH^\pm_\ell,
\qquad \HH^+_\ell\cong \HH^-_\ell \cong 
U^{(\ell+1)}\otimes \DD^{(\ell)}.
\eea
Let $E_\ell$ be the projections onto the subspaces
$\HH_\ell=\HH^+_\ell\oplus \HH^-_\ell$, $E^{(h)}=\sum_{\ell\geq h}E_\ell$, and
$\HH^{(h)}=E^{(h)}\HH=\bigoplus_{\ell\geq h}\HH_\ell$.
Let $P_\mu$, $M_{\mu\nu}$, $D$, $K_\mu$ the generators of the
conformal Lie algebra \eref{cfLA} represented on the one-particle space of the
complex massless Klein-Gordon field.

The main result defines deformed generators $\PP_k$ (the translations
of the deformed QFT) in terms of the generators of the scalar QFT on
the subspace $\HH^{(h)}$ of the one-particle space of the scalar QFT. 
The deformation is a simple ansatz for a vector operator of
  scaling dimension 1,  that (unlike the undeformed $P_k$) admits also transitions $\HH^\pm_\ell\to \HH^\pm_\ell$.

\begin{proposition}\label{p:main}
Let $h$ a non-negative integer. For $k=1,2,3$, make an ansatz for self-adjoint deformed
generators $\PP_k$, $\MM_{0k}$, $\KK_k$ of $\mm$ on $\HH^{(h)}=E^{(h)}\HH$ by 
\bea{PP}\PP_k:= \sum\nolimits_{\ell\geq h} a_\ell\cdot(E_{\ell+1}P_kE_\ell+E_\ell
P_kE_{\ell+1}) + \sum\nolimits_{\ell\geq h} b_\ell \cdot Q \cdot \eps_{kmn}P_0M_{mn} E_\ell,\eea
\bea{KK} 2\MM_{0k}:=i[K_0,\PP_k],\qquad  -\KK_k:=i[K_0,\MM_{0k}],\eea
where the coefficients $a_\ell$, $b_\ell$ are real. 

(i) The deformed generators $\PP_k$, $\MM_{0k}$, $\KK_k$ satisfy the correct commutation relations \eref{PCT} with the
PCT operator. \\[1mm]
(ii)
Together with the undeformed generators $P_0$, $D$, $K_0$ of $\mb$
and $M_{kl}$ of $\rr$, they satisfy the conformal Lie algebra \eref{cfLA} on
$\HH^{(h)}$ if and only if   
\bea{ab} 
a_\ell^2 = \frac{(\ell+1)^2-h^2}{(\ell+1)^2}, \qquad b_\ell^2 =
\frac{h^2}{4\ell^2(\ell+1)^2}
\eea
and all coefficients $b_\ell$ have the same sign. 
\\[1mm]
(iii) The generators $\PP_k$ as specified by (ii) 
satisfy the mass-shell condition on $\HH^{(h)}$:
\bea{MS}\sum\nolimits_k\PP_k^2=P_0^2.
\eea 
(iv) The resulting representation of the Lie algebra $\cf$ integrates
to a true (i.e., not a covering) representation $\UU$ of the conformal
group, which is equivalent to the irreducible representation $U_{(h)}$
of the conformal group.
\end{proposition}

Proof: (i) is immediate by \eref{QJ}.  To prove (ii), we start with a Lemma.  
\begin{lemma} (i) The deformed generators \eref{PP}, \eref{KK} fulfill
  the correct $[\hh,\mm]$ commutation relations \eref{vector} and
  \eref{module} independent of the
  specification of the coefficients. \\[1mm]  
(ii) The remaining $[\mm,\mm]$ commutation relations are also true on
$\HH_\ell$ ($\ell\geq h$), if
and only if 
\bea{PK}
i[\PP_k,\KK_l]=2\delta_{kl}D+2M_{kl}.
\eea
\end{lemma}
Proof of the Lemma: (i) is obvious because the projections $E_\ell$
and the charge operator $Q$ commute with $\hh=\mb\oplus\rr$, and the
remaining operators $P_k$ and $\eps_{kmn}M_{mn}$ transform like 
vectors under $\rr$ and commute with $P_0$ and have the correct
commutator with $D$ (``scaling dimension 1'',  \eref{cfLA}), while the
commutators with $K_0$ are part of the definition. \\[1mm]
(ii) follows by repeated application of $\ad_{K_0}$ and $\ad_{P_0}$ to \eref{PK}, using
\eref{module} and \eref{vector}. \qed

(Alternatively, 
the correct Poincar\'e commutation relations
$$i[\PP_k,\PP_l]=0,\quad i[\MM_{0k},\PP_l]=\delta_{kl}P_0, \quad
i[\MM_{0k},\MM_{0l}]=M_{kl}$$ would also serve the same purpose as \eref{PK}.)

The remaining task for the proof of (ii) of the proposition is just a
verification of \eref{PK}, where $\PP_k$ is given by \eref{PP}
  and
  $$\KK_l= \sum\nolimits_{\ell\geq h} a_\ell\cdot(E_{\ell+1}K_kE_\ell+E_\ell
P_kE_{\ell+1}) - \sum\nolimits_{\ell\geq h} b_\ell \cdot Q \cdot \eps_{kmn}K_0M_{mn}E_\ell
$$
by \eref{KK}. Using \eref{actmb}--\eref{actm}, one has to prove that
for harmonic homogeneous polynomials $Y_\ell$ of degree $\ell$, the
commutators $[\PP_k,\KK_l]$ make no transitions $\HH_\ell\to
\HH_{\ell\pm 2}$: 
$$E_{\ell\pm2}\big(\PP_kE_{\ell\pm1}\KK_l
-\KK_lE_{\ell\pm1}\PP_k\big)\ket{Y_\ell}^\pm_t   \,\stackrel!=\, 0,$$
(which turns out to be automatically satisfied by the ansatz), as well as
\bea{conds}E_{\ell\pm1}\big(\PP_k(E_{\ell\pm1}+E_\ell)\KK_l
-\KK_l(E_{\ell\pm1}+E_\ell)\PP_k\big)\ket{Y_\ell}^\pm_t  
&\stackrel!=& 0,\\ \notag E_{\ell}\big(\PP_k(E_{\ell+1}+E_\ell+E_{\ell-1})\KK_l
-\KK_l(E_{\ell+1}+E_\ell+E_{\ell-1})\PP_k\big)\ket{Y_\ell}^\pm_t  &\stackrel!=&
-2i(\delta_{kl}D+M_{kl})\ket{Y_\ell}^\pm_t,\eea
where the actions of the subsequent operators depend on the
intermediate projections, given by  \eref{deco-pY}. E.g., the first
term in the first condition \eref{conds} (for $\ell+1$) becomes
$$ia_\ell b_{\ell+1}Q\eps_{kmn}P_0M_{mn}E_{\ell+1}K_l\ket{Y_\ell}^\pm_t -
ia_\ell b_\ell E_{\ell+1}P_kQ\eps_{lmn}K_0M_{mn}\ket{Y_\ell}^\pm_t$$
which has to be worked out with \eref{actmb}--\eref{actm}.
Exploiting the identity for harmonic homogeneous
polynomials $Y_\ell$ (proven by contraction with $\eps_{akl}$ and
using \eref{deco-pY})
$$E_{\ell+1}\ket{(\eps_{lmn}p_k-\eps_{kmn}p_l)p_m\partial_nY_\ell}_t =
\ell\cdot\eps_{jkl} E_{\ell+1}\ket{p_jY_\ell},$$
one finds that the first condition is satisfied if and only if 
\bea{C1}
\ell \cdot a_\ell b_\ell = (\ell+2)\cdot a_\ell b_{\ell+1}.
\eea
Similarly, the second condition
\eref{conds} reduces to
\begin{subequations}
\bea{C2} a_\ell^2+4\ell^2\cdot b_\ell^2 &=& 1,\\ \label{C3}
 a_\ell^2-a_{\ell-1}^2 &=& 4(2\ell+1)\cdot b_\ell^2.\eea
\end{subequations}
Eliminating $b_\ell^2$ from \eref{C2} and \eref{C3}, one gets a
simple recursion
$$(\ell+1)^2a_\ell^2-\ell^2 a_{\ell-1}^2
= (\ell+1)^2-\ell^2,$$ hence $(\ell+1)^2(a_\ell^2-1)=const$.  
The initial condition $a_{h-1}=0$ gives $const=-h^2$, hence \eref{ab}.
\eref{C1} shows that $b_\ell$ have constant sign. 

This proves (ii). Evaluation of \eref{MS} on arbitrary vectors
$\ket{Y_\ell}^\pm_t$ with the given values \eref{ab}, yields the
desired result, proving (iii).

Finally, to prove (iv) we note that the spectrum of $L_0$ in the
deformed representation of $\cf$ is a subset of its spectrum on $\HH^{(0)}$,
hence integer. Therefore $\UU$ integrates to a true representation of
the conformal group. Knowing the multiplicities of the subrepresentations of
$\UU\vert_{\Mob\times\SO(3)}$, and comparing with \eref{repns}, one
concludes that $\UU$ is unitarily equivalent to the irreducible
representation $U_{(h)}$.
\qed

The signs of $b_\ell$ may be chosen positive without loss of
generality, via the unitary charge conjugation. Also the coefficients
$a_\ell$ may all be chosen positive via a unitary involution in the
center of $\mb\oplus\rr$.  

\subsection{Field algebras} 
The construction of a local QFT on the Fock space
$\FF^{(h)}=\Gamma(\HH^{(h)})$ over $\HH^{(h)}$ is routine. 

The deformed representation $\UU$ of the conformal group (equivalent to $U_{(h)}$ by \pref{p:main}(iv)) on the one-particle space $\HH^{(h)}$ lifts to the Fock space
$\FF^{(h)}$ by second quantization.

For open intervals $I\subset \RR$ (the time
axis), let $O_I$ be the corresponding doublecone spanned by $I$, and
define $\AA(O_I):= A(O_I)\vert_{\FF^{(h)}}$, where $A(O)$ are the local
algebras of the scalar field theory. For arbitrary doublecones, choose
an interval $I$ and a conformal transformation $g$ such that $gO_I=O$,
and define 
$$\AA(O):=\UU(g)\AA(O_I)\UU(g)^*.$$ 
The definition is unambiguous because if $g_1O_{I_1}=O= g_2O_{I_2}$, then
$g_2^{-1} g_1(I_1)=I_2$, hence $g:=g_2^{-1} g_1\in \Mob\times\SO(3)$,
hence $\UU(g)A(I_1)\UU(g)^*=A(I_2)$. Thus, the net 
$$O\mapsto\AA(O)$$ is
conformally covariant.

Because for any pair of spacelike separated
doublecones, there is a conformal transformation $g$ mapping the
doublecones into the forward and backward lightcones, 
respectively, locality follows from the Huygens property of the
scalar field ($A(V_+)$ commutes with $A(V_-)$) by covariance
(see, e.g., \cite{LMR}).  

\subsection{Field equations}

At the level of fields, we define a symmetric and traceless field tensor
$F^{(*)}_{j_1\dots j_h}(t,\vec 0)$  by identification with
the multiplet of derivative fields $Y_{j_1\dots
  j_h}(-i\vec\nabla)\varphi^{(*)}(t,\vec 0)$ (where $Y_{j_1\dots
  j_h}$ is the appropriately normalized symmetric traceless tensor of harmonic
polynomials of spin $\ell=h$) restricted to $\FF^{(h)}$, and hence $F^{(*)}_{j_1\dots j_h}(t,\vec 0)\Omega$ with the
(improper) spin-$h$ vectors $\ket{Y_{j_1\dots j_h}}^{\pm}_t$ of
$\HH^\pm_h$. 
We define $F^{(*)}_{j_1\dots j_h}(t,\vec x)$ by the adjoint action of 
$\UU(\vec x)=e^{i x_k\PP_k}$ such that $F^{(*)}_{j_1\dots j_h}(t,\vec x)\Omega$
are (improper)
vectors in $\HH^{(h)\pm}$. Then $F^{(*)}(x)$ smeared within
a doublecone $O$ are affiliated with the algebra $\AA(O)$, hence they
are local fields.

In order to make the identification of the deformed field thus
  defined with the
free field of helicity $\pm h$, we have to establish the equation of
motion 
\cite{LMPR} 
\bea{HMax1}
\nabla_kF_{kj_2\dots j_h} = 0,\qquad \eps_{kjm} \nabla_{k} F_{jj_2\dots j_h} = i\partial_t
F_{mj_2\dots j_h},
\eea
where $F=E+iB$, and $E_{j_1\dots j_h}$ and $B_{j_1\dots j_h}$
are symmetric traceless ``electric'' and ``magnetic'' tensors. On the
time axis, we have by construction 
$$F_{j_1\dots j_h}(t,\vec 0)\Omega =  \ket{Y_{j_1\dots
  j_h}}^+_t,\quad\hbox{hence}\quad \nabla_kF_{j_1\dots j_h}(t,\vec
0)\Omega = i\PP_k\ket{Y_{j_1\dots j_h}}^+_t.$$
With \pref{p:main}, we compute 
\bea{PPh}
\PP_k\ket{Y_{j_1\dots j_h}}^+_t =a_hE_{h+1}\ket{p_kY_{j_1\dots j_h}}^+_t
-2b_h\cdot \sum\nolimits_a\eps_{kmn}\partial_t\ket {p_m\partial_nY_{j_1\dots j_h}}^+_t. 
\eea
With \eref{deco-pY}, this implies 
\bea{HMax2}\PP_k\ket{Y_{kj_2\dots j_h}}^+_t=0, \qquad \eps_{kjm}
\PP_k\ket{Y_{jj_2\dots j_h}}^+_t = -2b_h\partial_t \ket{(p_n\partial_m-p_m\partial_n)Y_{nj_2\dots j_h}}^+_t, \eea
because $\big(p_k-\frac{\vec p\,^2}{2h+1}\partial_k\big) Y_{kj_2\dots
  j_h}$ and $\eps_{kmn}p_m\partial_n Y_{kj_2\dots
  j_h}$ and $\eps_{kjm} \big(p_k-\frac{\vec p\,^2}{2h+1}\partial_k\big) Y_{jj_2\dots
  j_h}$ are symmetric traceless tensors of rank $r=h-1,h-1,h$, respectively, of harmonic
homogeneous polynomials of degree $\ell = h+1,h,h+1$, respectively,
which vanish since $r\neq\ell$. 

Now, the recursion $Y_{nj_2\dots j_h}= \big(p_n-
  \frac{\vec p\,^2}{2h-1}\partial_n\big)Y_{j_2\dots j_h}$ together
  with harmonicity and homogeneity of $Y_{j_2\dots j_h}$ imply 
$(p_n\partial_m-p_m\partial_n)Y_{nj_2\dots j_h} =
-(h+1)Y_{mj_2\dots j_h}$. Since  $2(h+1)b_h=1$, we conclude that the
higher Maxwell equations hold on the time axis and on the vacuum vector: 
\bea{HMax0}
\nabla_kF_{kj_2\dots j_h}(t,\vec 0)\Omega = 0,\qquad \eps_{kjm} \nabla_{k} F_{jj_2\dots j_h}(t,\vec 0)\Omega = i\partial_t
F_{mj_2\dots j_h}(t,\vec 0)\Omega.\eea
The complex conjugate higher
Maxwell equations for $F^*=E-iB$ are guaranteed by the presence of the operator $Q$
in \eref{PP}, that switches the sign of $i$ in the right-hand side of
\eref{PPh} for the vectors $\ket {Y}_t^-$ of charge $-1$.

At this point, it becomes apparent how the charge of the scalar field
is re-interpreted as the sign of the helicity of the higher Maxwell field.

By applying the spacelike translations $\UU(\vec x)$ to \eref{HMax0},
we conclude that the higher Maxwell equations on the vacuum vector hold
everywhere in Minkowski space. Because $F^{(*)}$ are local fields on
the time axis, by conformal covariance they are local on Minkowski
spacetime. Then the
Reeh-Schlieder theorem ensures that the higher Maxwell equations hold
as operator equations.

\subsection{Unitary implementation}
\label{s:uni}

It is clear from our identification of the helicity-deformed
representation on the subspace $E^{(h)}\HH_0$  of
the one-particle space for the complex
scalar field with the known representation on the higher-helicity
one-particle space, that there
must exist a unitary operator
$U$ intertwining the two representations. While our result \pref{p:main} was found without knowing
this unitary, we could identify it {\em a posteriori} for the case
$h=1$ (Maxwell). We sketch the essential steps.

To prevent confusion, we rename the one-particle Hilbert space and its
(improper) vectors of \sref{s:main} as $\HHs$ and $\kets Y_t{}^\pm=
Y(-i\vec \nabla)\varphi^{(*)}(x)\Omega\vert_{x=(t,\vec0)}$, and we
introduce the one-particle Hilbert space of the
Maxwell theory $\HHm$ and its (improper) vectors
$$\ketm{\vec Y}_t{}^\pm:= \frac1{\sqrt2}\cdot Y_j(-i\vec\nabla)\big(E_j(x)\pm iB_j(x)\big)\Omega\vert_{x=(t,\vec0)},$$
where $\vec Y=(Y_1,Y_2,Y_3)$ is a triple of polynomials, and the sum convention
for vector indices is understood. We consider only the case
of charge $+$ and suppress the superscript. The case of charge $-$ is
identical, up to a change of sign, see below.

The respective inner products deriving from the respective two-point
functions are
$${}^{\rm scl}\!\!{}_t\braket Y{Y'}^{\rm scl}_{t'}=(2\pi)^{-3}\int
\frac{d^3p}{2p^0}\,\ol{Y(\vec p)}Y'(\vec p)\,
e^{-i(t-t')} \quad (p_0=\vert\vec p\vert),$$
$${}^{\rm Maxw}\!\!{}_t\braket {\vec Y}{\vec Y'}^{\rm Maxw}_{t'}=(2\pi)^{-3}\int
\frac{d^3p}{2p^0}\,\ol{Y_i(\vec p)}\big(\delta_{ij}p_0^2-p_ip_j +i\eps_{ijk} p_0p_k\big)Y_j'(\vec p)\, e^{-i(t-t')}.$$
For the conjugate field $\vec E-i\vec B$, only the $\eps_{ijk}$-term
would change sign. We write the latter as
\bea{MTS}{}^{\rm Maxw}\!\!{}_t\braket {\vec Y}{\vec Y'}^{\rm Maxw}_{t'}
= {}^{\rm scl}\!\!{}_t\bra{Y_i}T_{ij}\ket{Y_j'}^{\rm scl}_{t'}
\eea
where
$$T_{ij}=\delta_{ij}\vec P_0^2-P_iP_j +i\eps_{ijk} P_0P_k.$$
The null vectors of these inner products are \eref{waveY} respectively
$$\ketm{\vec p\,^2\,\vec Y}_t+\partial_t^2\ketm{\vec Y}_t=0,\quad \ketm{\vec
  p\,Y}_t=0,\quad \ketm{\vec p\times\vec Y}_t+\partial_t\ketm{\vec Y}_t=0.$$

The action of the conformal generators on the vectors in $\HHm$ is
similar as \eref{actmb}--\eref{actm}, with modifications due to the vector
character of the field and its scaling dimension 2. They will be
displayed in due context in the proof of \pref{p:uni}.

We now introduce operators $V_i$ and $\nabla_i$ on $\HHs$:
\bea{Vi}V_i:=M_{ki}P_k-\frac i2P_0\eps_{ikl}M_{kl},\qquad \nabla_i \kets Y_t
:= \kets{\partial_i Y}_t
\eea
and operators $V:\HHm\to\HHs$ and $V^*:\HHs\to\HHm$:
\bea{V}
V\ketm{\vec Y}_t :=V_i\kets{Y_i}_t,\qquad V^*\kets Y_t:=
-i\ketm{\vec\nabla Y}_t.
\eea
Here, $\vec\nabla Y(\vec p)$ denotes the gradient w.r.t.\ $\vec p$. These
operators are well-defined because they respect the respective
null vectors.

The following Lemma and Proposition provide the desired unitary operator.
\begin{lemma}
  \label{l:uni} (i) The range of $V_i$ is orthogonal to the $\ell=0$ subspace
  $E_{0}\HHs$. \\[1mm]
  (ii) It holds $T_{ij}\nabla_j = iV_i^*$ \\[1mm]
  (iii) It holds $T_{ij}=V_i^*C^{-1}V_j$ where $C=\frac12 M_{kl}M_{kl} =
  \sum_\ell\ell(\ell+1) E_\ell$ is the Casimir operator.\\[1mm]
  (iv) $V^*$ is the adjoint of $V$. 
  \end{lemma}

  \begin{proposition} \label{p:uni} (i) It holds $VV^*=C$. \\[1mm]
    (ii) The
    operator $U=C^{-\frac12}V:\HHm\to (\eins-E_{0})\HHs$ is
    unitary. In other words, $U$ arises by polar decomposition of
    $V^*=U^*C^{\frac12}$. \\[1mm]
    (iii) $U$ intertwines the actions of
    $\mb\oplus\rr$ on $\HHm$ and $(\eins-E_{0})\HHs$.\\[1mm]
    (iv) $U$
    intertwines the actions of the generators $P_k$, $M_{0k}$, $K_k$
    of $\mm$
    on $\HHm$ with the actions of the deformed generators of $\mm$
    according to \pref{p:main} on
    $(\eins-E_{0})\HHs$. 
  \end{proposition}

  The proofs proceed by direct computation heavily using
  \eref{actmb}--\eref{actm} and the present definitions. It could be rewarding to
  have a more elegant proof providing 
a better insight into the algebraic nature of the deformation.

  Proof of the Lemma: (i) Obvious because $M_{kl}E_{0}=0$. \\[1mm]
  (ii) Straight computation, using \eref{actso}.\\[1mm]
  (iii) The
  lengthy computation goes as follows: One first shows, for
  harmonic polynomials $Y_\ell$ of degree $\ell$, that $\nabla_kC^{-1}V_j\kets {Y_\ell}_t =
  i\delta_{jk}\kets {Y_\ell}_t + $ terms annihilated by $T_{ik}$ (summation over
  $k$), and then uses (ii). For the first step, the identities
  $$E_{\ell+1} P_iM_{ij}E_\ell= i\ell\cdot E_{\ell+1}P_jE_\ell,\quad 
E_{\ell+1} M_{ij}P_iE_\ell= i(\ell+2)\cdot E_{\ell+1}P_jE_\ell$$
and their  adjoint are extensively exploited. \\[1mm]
(iv) Straight computation:
$${}^{\rm Maxw}\!\!{}_t\bra{\vec Y}V^*\kets{Y'}_{t'} 
\,\stackrel{\rm Def}=\, -i\,{}^{\rm Maxw}\!\!{}_t\braket{\vec
  Y}{\vec\nabla Y'}^{\rm Maxw}_{t'} 
= -i\,{}^{\rm scl}\!\!{}_t\bra{\vec Y}T_{ij}\nabla_j \kets{Y'}_{t'} = $$
$$\,\stackrel{\rm (ii)}=\, {}^{\rm scl}\!\!{}_t\bra{Y_i}V_i^*\kets{Y'}_{t'}
=\ol{{}^{\rm scl}_{\hskip2mm t'}\bra{Y'}V_i\kets{Y_i}_{t}} 
\,\stackrel{\rm Def}=\,\ol{{}^{\rm scl}_{\hskip2mm t'}\bra{Y'}V\ketm{\vec Y}_{t}}.
$$
\qed

  Proof of the Proposition: (i) Straight computation: $VV^*
  =-iV_i\nabla_i = C$ because the $i\eps_{ikl}$-terms vanish. \\[1mm]
  (ii) On the range $(\eins-E_0)\HHs$ of $V$, $C^{-\frac12}$ is well
defined. Then $UU^* = C^{-\frac12}VV^*C^{-\frac12}=\eins$ by virtue of
(i). On
the other hand, using (ii) of the Lemma,
$${}^{\rm Maxw}\!\!{}_t\bra{\vec Y}U^*U\ketm{\vec Y'}_{t'}
={}^{\rm scl}\!\!{}_t\bra{Y_i}V_i^*C^{-1}V_j\kets{Y_j'}_{t'}={}^{\rm scl}\!\!{}_t\bra{Y_i}T_{ij}\kets{Y_j'}_{t'}
={}^{\rm Maxw}\!\!{}_t\braket{\vec Y}{\vec Y'}^{\rm Maxw}_{t'},
$$
hence $U^*U=\eins$ as well. \\[1mm]
(iii) Because $M_{kl}$ commute with $C$ and $\vec V$ transforms as a
vector, it is easily seen that the chain of
maps $U^*P_0U$ takes $\ketm{\vec Y}_t$ to
$-i\ketm{(p_k\partial_l-p_l\partial_k)\vec Q+Q_l\vec e_k-Q_k\vec e_l}_t$, which is
the action of  $M_{kl}$ in the Maxwell one-particle space; the
additional terms as compared to \eref{actso} account for the vector
character of the Maxwell field.  
\\[1mm]
Because $P_0$ commute with $V_i$ and with $C$, $U^*P_0U $ takes 
$\ketm{\vec Y}_t$ to $-i\partial_t\ketm{\vec Y}_t=P_0\ketm{\vec
  Y_t}$. \\[1mm]
Because $V_i$ scale like $\vec P$, $U^*DU$ takes
$\ketm{\vec Y}_t$ to $-it\partial_t\ketm{\vec Y}_t-i
\ketm{(\vec p\cdot\vec \nabla)\vec Y+2\vec Y}_t$, which is the action of $D$ in the
Maxwell one-particle space. The factor $2$ as compared to \eref{actmb}
reflects to the scaling dimension $2$ of the Maxwell field.
\\[1mm]
Finally, for $K_0$, it suffices to note that the equality of Casimir operators
$\frac12M_{kl}M_{kl}=C=\frac12(P_0K_0+K_0P_0)-D^2=P_0K_0-D^2-iD$ 
holds in both representations by \eref{repns}. Since $U$ intertwines $M_{kl}$,
$D$ and $P_0$, it also intertwines $C$ and $K_0=P_0^{-1}(C+D^2+iD)$.
\\[1mm]
(iv) It suffices to verify $UP_kU^*=\widetilde P_k$. The others then  
follow by \eref{module} because $U$ intertwines $K_0$. One 
computes the various contributions $E_{\ell'}VP_kV^*\kets{Q_\ell}_t$,
and by dividing by the respective eigenvalues of $C^{\frac12}$, one
obtains \eref{PP} with the correct coefficients \eref{ab}.
\qed

\section{Mass deformation}\label{s:mass}
\setcounter{equation}{0}
A second, and much simpler, instance of spacelike deformation is the
construction of the massive Klein-Gordon field as a deformation of the
spacelike translations and boosts of the massless Klein-Gordon
field. It is ``complementary'' to the corresponding Hamiltonian deformation 
treated in \cite{EF}.

In the present instance, we can just write down the deformed
generators. Because
the massive one-particle vectors have energy $p^0\geq m$, the spectral projection
$E_m=\theta(P_0-m)$ will play the role of the projection $E^{(h)}$ in
\sref{s:main}.   

The Lie algebra of the Poincar\'e group again has a
symmetric space decomposition 
$\hh\oplus\mm$, where $\hh=\t\oplus\rr$ ($\t=$ time
translations), and $\mm$ is spanned by the spacelike momenta and the
boost generators. By definition, the rotations $M_{kl}$ and the
Hamiltonian $P_0$ remain undeformed. Thus, the deformed boosts
$\MM_{0k}$ also determine the deformed momenta $\PP_k=-i[P_0,\MM_{0k}]$. 

We return to the real scalar field, and denote by $\varphi_0$ and $\varphi_m$
the massless and massive fields.

For the massless field, we write
$\ket{Y}_t:=Y(-i\vec\nabla)\varphi_0(t,\vec0)\Omega$ as 
before, and for the massive field denote
$\ket{Y}^m_t:=Y(-i\vec\nabla)\varphi_m(t,\vec0)\Omega$. The massive two-point function yields
\bea{spher}
{}^m\braket{Y,f}{Y',f'}^m = (2\pi)^{-3}\int
\frac{p^2\,dp}{2p^0}\cdot \ol{\wh f(p^0)}\wh f'(p^0)
\int d\sigma(\vec n) \, \ol{Y(p\vec n)}Y'(p\vec n),
\eea
where
$p^0=\sqrt{p^2+m^2}$, $\vec p=p\vec n$, and $d\sigma$ is the invariant
measure on the unit sphere. The Klein-Gordon equation becomes
$\ket {\vec p\,^2Y}_t^m +(\partial_t^2+m^2)\ket{Y}^m_t=0$ w.r.t.\ the
inner product \eref{spher}.

The inner product \eref{spher} is diagonal w.r.t.\ the spin of the polynomials $Y$,
$Y'$. Consequently, it is a direct sum of inner products  for
polynomials $Y_\ell$ of spin $\ell$ as before 
\bea{spher-l}
{}^m\braket{Y_\ell,f}{Y_\ell',f'}^m = (2\pi)^{-3}\int
\frac{p^{2\ell+2}\,dp}{2p^0}\cdot p^{2\ell}\cdot \ol{\wh f(p^0)}\wh f'(p^0)
\int d\sigma(\vec n)\, \ol{Y_\ell(\vec n)}Y_\ell'(\vec n),\quad
\eea
Passing to the integration
variable $\omega=p_0$, one has 
$$\frac{p^{2\ell+2}dp}{2p^0} =
\frac12(\omega^2-m^2)^{\ell+1/2}d\omega.$$
It will be advantageous to  pass to their Fourier transforms
$\ket{Y}^m_\omega = \int dt\,e^{-i\omega t} \,
\ket{Y}^m_t$, such that
\bea{massY}
\ket {\vec p\,^2Y}_\omega^m =(\omega^2-m^2)\ket{Y}^m_\omega.
\eea
Then
the identification $U:\HH_m\to E_m\HH_0$ of the massive
one-particle space $\HH_m$ with the subspace $E_m \HH_{0}$ of the
massless one-particle space $\HH_0$
\bea{Uell}
U:\ket{Y_\ell}^m_\omega \mapsto \sigma(\omega)^{\ell+\frac12}\ket{Y_\ell}_\omega,\quad\hbox{where}\quad \sigma(\omega) = \Big(1-\frac{m^2}{\omega^2}\Big)^{\frac12},
\eea
is unitary w.r.t.\ the respective inner products \eref{spher} for
$m=0$ and for $m>0$. In view of \eref{spher}, the same equation
\eref{Uell} remains true for all homogeneous polynomials of degree
$\ell$; this 
is compatible with \eref{waveY} and \eref{massY} by virtue of
$\omega^2-m^2 = \omega^2\cdot \sigma(\omega^2)$. 

The deformed Poincar\'e generators on $E_m\HH_{0}$ arise by the
unitary conjugation $\Ad_U$ of the known action of the massive
Poincar\'e generators on $\HH_m$, i.e., the pull-back under the
identification \eref{Uell}. 
The massive Poincar\'e generators $P_\mu$ and $M_{\mu\nu}$ act
on $\ket{Y_\ell}^m_t$ in exactly the same way as the corresponding massless
generators on $\ket{Y_\ell}_t$ in \eref{actmb}--\eref{actm}. In particular,
the deformation preserves the Hamiltonian $P_0$ and the generators
$M_{kl}$ of rotations. The deformation of the spacelike momenta gives immediately 
\bea{Pmass}
\PP_k\ket{Y_\ell}_\omega=\sigma(\omega)\cdot P_k\ket{Y_\ell}_\omega\quad\Rightarrow\quad
\PP_k=P_k\cdot\Big(1-\frac{m^2}{P_0^2}\Big)^{\frac12}.
\eea
The mass-shell condition
$$\sum\nolimits_k\PP_k^2=P_0^2-m^2$$
is trivially fulfilled by \eref{Pmass}. 

For the deformed boosts on
$E_m\HH_0$,
one gets 
\bea{M}
\notag
\MM_{0k}\ket{Y_\ell}_\omega&=&\sigma(\omega)^{-\ell-\frac12}\Big(\partial_\omega
\big(\sigma(\omega)^{\ell+\frac32}\ket{p_kY_\ell}_\omega\big) +
\omega\sigma(\omega)^{\ell-\frac12}\ket{\partial_k
Y_\ell}_\omega\Big)\\\notag
&=&\big((\ell+\frac32)\sigma'(\omega)+\sigma(\omega)\partial_\omega
\big)\ket{p_kY_\ell}_\omega+
\omega\sigma(\omega)^{-1}\ket{\partial_kY_\ell}_\omega.\eea
Using \eref{actmb}--\eref{actm}, this can be seen to be equivalent to
\bea{MD} 
\MM_{0k}=
\Big(M_{0k}-\frac1{2P_0}(DP_k+P_kD)\cdot\frac{m^2}{P_0^2}\Big)\cdot\Big(1-\frac{m^2}{P_0^2}\Big)^{-\frac12}
\eea
(where the operator ordering has been adjusted so as to match the
coefficient $\ell+\frac32$). Because in the
massless one-particle representation the Casimir operators $C=\frac12
M_{kl}M_{kl}$ of $\rr$ and $\frac12(P_0K_0+K_0P_0)-D^2$ of $\mb$
coincide, this can also be written
\bea{MC}\MM_{0k}= \frac12\Big(\sigma(P_0)M_{0k}+M_{0k}\sigma(P_0)\Big) +
\frac i2\,[C,P_k]\cdot\sigma'(P_0).
\eea
By construction (unitary conjugation with $U$) the deformed generators
on the subspace $E_m\HH_{0}$ are self-adjoint and satisfy the
Poincar\'e commutation relations. Indeed, the hermiticity, as well as the commutator
$i[P_0,\MM_{0k}]=-\PP_k$, are also {\em explicitly} verified without much effort. 
The {\em explicit} verification of the commutation relation
$i[\MM_{0k},\MM_{0l}]=M_{kl}$ does not directly give the desired
result, but rather
$$i[\MM_{0k},\MM_{0l}] =
\Big(M_{kl}+\big(M_{0k}P_l-M_{0l}P_k)\frac{m^2}{P_0^3}\Big)\Big(1-\frac{m^2}{P_0^2}\Big)^{-1}.$$
This equals $M_{kl}$ on $E_m\HH_0$ because
$M_{0k}P_l+M_{l0}P_k+M_{kl}P_0=\eps_{klj}W^j$, and the Pauli-Lubanski operator 
$W^\mu=\frac12\eps^{\mu\nu\kappa\lambda}
M_{\nu\kappa}P_\lambda$ vanishes in the massless scalar representation.

\section{Summary}

We have presented 
two families of examples of spacelike deformations that
allow to construct new quantum field theories by fixing the restriction 
of a given QFT to the time axis, and deforming only the ``transverse''
symmetry generators. The remarkable feature is that the scheme admits
the change of discrete quantum numbers (the helicity in our first
example). 

Both instances of spacelike deformation presented here make essential
use of the enveloping algebra of the Lie algebra of the respective
spacetime symmetry group (conformal, resp.\ Poincar\'e). 

In both cases, it is true that we knew the expected deformation from
the outset. But only in the mass deformation case did we know the unitary
  operator that transfers the massive generators to the massless
  one-particle space, and we used this
knowledge to compute the deformed generators. In contrast, \pref{p:main} is a uniqueness result, once the
subspace is specified on which the deformation is supposed to be
defined.

It is not difficult to see that one can also deform any given helicity
$h'>0$ to a helicity $h>h'$,  
and any given mass $m'>0$ to a mass $m>m'$, as would be expected
from the underlying pattern of inclusions of Hilbert spaces. One only
needs to re-adjust the numerical coefficients of $E_{\ell'}\PP_kE_{\ell}$ in
\eref{PP}, and replace $\sigma(P_0)$ in \eref{MC} by
$\sqrt{(P_0^2-m^2)/(P_0^2-m'^2)}$.
On the other hand,
increasing mass and spin simultaneously might not be possible by lack
of an inclusion of one-particle representations of the subgroup fixing
the time axis. However, it seems possible to get a
  representation with $m>0$ and $s>0$ by deforming the direct sum of
  massless helicity representations running over all $\vert h\vert\leq s$,
  because conversely, the restriction of a massive spin-$s$
    representation to the rotation subgroup is the direct sum of the
    restrictions of helicity-$h$ representations with $\vert h\vert\leq
    s$.

The case of interacting theories will need methods going beyond
representation theory of spacetime symmetry groups. Similar ideas
leading to integrable models in two spacetime dimensions were
previously pursued in \cite[Sect.\ 4]{BT}. Here, based on an operator-algebraic deformation
result for chiral conformal QFT \cite{LW}, models with translation 
generators $P_0= \frac12(P+ m^2/P)$ and $P_1=\frac12(P-m^2/P)$,
satisfying the mass-shell condition $P_0^2-P_1^2=m^2$, were constructed starting from a Möbius covariant QFTs with generator $P$.

\section{Outlook}

Our constructions may give insights into the modular theory of
local algebras for massive theories \cite{FG,Sa}, which is not as well known as for
massless theories. Let us explain the idea.

In a generic QFT, if the local algebras $A(O_I)=A(I)$ for doublecones
$O_I$ spanned by an interval $I$ along the time axis are given, then
they are defined for general doublecones by the adjoint action of Poincar\'e
transformations. In \sref{s:main} and \sref{s:mass}, the deformed
local algebras on the time axis arise just by restriction of the
undeformed local algebras to the respective second-quantized subspace
$\Gamma(E^{(h)})\FF_0$ or $\Gamma(E_m)\FF_0$.   

In the case of helicity deformations, one may adopt a different point
of view, referring only to the representations of the conformal group. 
Namely, given a unitary representation $\UU$ of $\Mob$ on $\HH$ and its
extension by the anti-unitary PCT operator $J$, the Brunetti-Guido-Longo construction 
\cite{BGL} (BGL) allows to define a real Hilbert space $H(\RR_+)\subset \HH$ 
such that $H(\RR_+)\cap iH(\RR_+)=\{0\}$ and $\overline{\CC\,
  H(\RR_+)}=\HH$. This definition uses only the dilations and
$J$. Acting with $\UU(g)$, $g\in\Mob$, one obtains a net of  
real standard subspaces $I\mapsto H(I)$ on the intervals of the circle.
This net of subspaces is local in the sense that the symplectic
complement $H(I)'\equiv (iH(I))^\perp$ of $H(I)$ coincides with
$H(I')$, where $I'$ is the complement of $I$ in $S^1$ and
orthogonality $\perp$ refers to the real part of the scalar
product. Upon second quantization, these properties turn into locality
of a M\"obius covariant chiral net of local 
algebras with the Reeh-Schlieder property. It trivially restricts to
a net on the time axis by deleting the point $-1\in S^1$ and identifying
$S^1\backslash\{-1\}$ with $\RR$ via the Cayley transform. 

If the unitary representation of $\Mob$ extends to a representation of the
four-dimensional conformal group on $\HH$, then the net
  of standard subspaces on $S^1$ extends to a
conformally covariant net $O\mapsto H(O)$ on the four-dimensional
Dirac manifold, which in turn restricts to a net on Minkowski
spacetime. By second quantization, one obtains a Huygens 
local net of local algebras $O\mapsto A(O)$. ``Huygens locality'' (=  
commutativity also at timelike distance) is a consequence of the locality
along the time axis, that is guaranteed by the BGL construction. 

By construction, the modular group of $H(\RR_+)$ is given by the
dilations, and that of $H(I)$ is the one-parameter subgroup of
$\Mob$ that fixes the interval $I$. It follows that the modular groups
of the local algebras $A(O)$ ($O$ a doublecone or a wedge) in the vacuum
state are the 
subgroups of the conformal group (conjugate to boost subgroups) that
fix the doublecone or wedge $O$. In the construction of \sref{s:main}, the
one-particle space is given by $E^{(h)}\HH$, the representation of the
M\"obius group remains undeformed, and the local subspaces and local
algebras away from the time axis are constructed with the deformed
translations and boosts. Because the projection $E^{(h)}$ commutes
with the representation of $\Mob$, it is automatic that the modular
groups on the time axis coincide with those of the scalar field
restricted to $E^{(h)}\HH$, and away from the time axis are conjugate
by deformed Poincar\'e transformations.

The situation is very different in the mass deformation of
\sref{s:mass}. Because the spectral projection $E_m$ does not commute with
the dilations, the latter are not defined on the subspace
$\HH_m$, and the BGL construction is not possible. Indeed, it is
well-known that in the massive case, $H_m(\RR_+)$ (to be identified
with $H_m(V_+)$ in the net on Minkowski spacetime) has trivial symplectic
complement (\cite{MT,SW}), in contrast to the duality
$(iH_0(\RR_+))^\perp=H_0(\RR_-)$ in the massless case. On the other hand,
we know that the massive local 
subspace $H_m(I)$ of an interval $I$ on the time axis coincides with the local
subspace $H_m(O_I)$ for the doublecone $O_I$ spanned by $I$; and by the
work \cite{EF} of Eckmann and Fr\"ohlich, we have a local unitary
equivalence between the massive and massless time-zero
algebras. Specifically, there is a unitary operator $U_R$ such that 
for intervals $I_r=(-r,r)\subset \RR$ symmetric around $t=0$ and $r<R$,
one has $H_m(O_r)=U_RH_0(O_r)$ where $O_r$ is the causal completion of
the time-zero ball of radius $r$. Thus, the modular groups of $H_m(O_r)$
are, for $r<R$, conjugate to the known modular groups of $H_0(O_r)$ by
$U_R$. Increasing $R$, the unitary $U_R$ will change, but the
subspaces $H(I_r)$ for $r<R$ and their modular groups remain
unchanged. Thus, the modular groups for $r<R_1<R_2$ commute with
$U_{R_2}U_{R_1}^*$, and a more detailed investigation of the unitaries
$U_R$ would be worthwhile to get a first insight into the hitherto
unknown massive modular groups. 

This information about the modular groups then passes to arbitary
doublecones via the adjoint action of the deformed translations and
boosts, as constructed in \sref{s:mass}.

\bigskip

\textbf{Acknowledgment.} We thank Roberto Longo for valuable
discussions, and the referee for suggesting improvements. Our work was supported in part by GNAMPA-INdAM, the MIUR
Excellence  
  Department Project awarded to the Department of Mathematics,
  University of Rome Tor Vergata, CUP E83C18000100006, and by the ERC Advanced Grant 669240 QUEST ``Quantum
  Algebraic Structures and Models''.

\end{document}